\documentclass[twocolumn,superscriptaddress,aps,preprintnumbers,amsmath,amssymb]{revtex4}
\usepackage{graphicx}% Include figure files
\usepackage{dcolumn}% Align table columns on decimal point
\usepackage{bm}% bold math

\begin{document}
%%%%%%%%%%%%%%%%%%%%%%%%%%%%%%%%%%%%%%
\def\a{\alpha}
\def\b{\beta}
\def\c{\varepsilon}
\def\d{\delta}
\def\e{\epsilon}
\def\f{\phi}
\def\g{\gamma}
\def\h{\theta}
\def\k{\kappa}
\def\l{\lambda}
\def\m{\mu}
\def\n{\nu}
\def\p{\psi}
\def\q{\partial}
\def\r{\rho}
\def\s{\sigma}
\def\t{\tau}
\def\u{\upsilon}
\def\v{\varphi}
\def\w{\omega}
\def\x{\xi}
\def\y{\eta}
\def\z{\zeta}
\def\D{\Delta}
\def\G{\Gamma}
\def\H{\Theta}
\def\L{\Lambda}
\def\F{\Phi}
\def\P{\Psi}
\def\S{\Sigma}

\def\o{\over}
\def\beq{\begin{eqnarray}}
\def\eeq{\end{eqnarray}}
\newcommand{\gsim}{ \mathop{}_{\textstyle \sim}^{\textstyle >} }
\newcommand{\lsim}{ \mathop{}_{\textstyle \sim}^{\textstyle <} }
\newcommand{\vev}[1]{ \left\langle {#1} \right\rangle }
\newcommand{\bra}[1]{ \langle {#1} | }
\newcommand{\ket}[1]{ | {#1} \rangle }
\newcommand{\EV}{ {\rm eV} }
\newcommand{\KEV}{ {\rm keV} }
\newcommand{\MEV}{ {\rm MeV} }
\newcommand{\GEV}{ {\rm GeV} }
\newcommand{\TEV}{ {\rm TeV} }
\def\diag{\mathop{\rm diag}\nolimits}
\def\Spin{\mathop{\rm Spin}}
\def\SO{\mathop{\rm SO}}
\def\O{\mathop{\rm O}}
\def\SU{\mathop{\rm SU}}
\def\U{\mathop{\rm U}}
\def\Sp{\mathop{\rm Sp}}
\def\SL{\mathop{\rm SL}}
\def\tr{\mathop{\rm tr}}

\def\IJMP{Int.~J.~Mod.~Phys. }
\def\MPL{Mod.~Phys.~Lett. }
\def\NP{Nucl.~Phys. }
\def\PL{Phys.~Lett. }
\def\PR{Phys.~Rev. }
\def\PRL{Phys.~Rev.~Lett. }
\def\PTP{Prog.~Theor.~Phys. }
\def\ZP{Z.~Phys. }

%%%%%%%%%%%%%%%%%%%%%%%%%%
\preprint{LAUR-07-7160}
\preprint{SLAC-PUB-12967}

\title{Minimal Direct Gauge Mediation}% Force line breaks with \\

\author{Masahiro Ibe}
\affiliation{%
Stanford Linear Accelerator Center, Stanford University, Stanford, CA 94309 and\\
Physics Department, Stanford University, Stanford, CA 94305
}%
%\altaffiliation[Also at ]{Physics Department, XYZ University.}
% \altaffiliation[Also at ]{Physics Department, XYZ University.}
\author{Ryuichiro Kitano}%
\affiliation{Theoretical Division T-8, Los Alamos National Laboratory, Los Alamos, NM 87545}
%\affiliation{%

%\affiliation{


\begin{abstract}
We propose a simple model of gauge mediation 
where supersymmetry is broken by a strong dynamics at $O(100)$\,TeV.
\end{abstract}

\maketitle
\subsection{Introduction}
One of the simplest and the most natural scenarios for supersymmetry breaking is to assume dynamical supersymmetry breaking at an energy scale of $O$(100)\,TeV.
The electroweak scale comes out as a one-loop factor lower scale than $O$(100)\,TeV 
via gauge mediation~\cite{Dine:1981za,Dine:1981gu,Dimopoulos:1982gm,
Affleck:1984xz,Dine:1993yw,Dine:1994vc,Dine:1995ag}. 
The scenario contains only a single scale as oppose to other scenarios where the supersymmetry breaking scale and the messenger scale are generated by different mechanisms. 
However, it has been known that a concrete model building of such a one-scale scenario 
is theoretically challenging (see Refs.~\cite{Izawa:1997hu,Izawa:1997gs,Nomura:1997uu,Izawa:1999vc,Izawa:2005yf,Murayama:2006yf,Ibe:2007wp,Nakayama:2007cf} 
for earlier attempts).
In this letter, we propose a simple (and possibly the simplest) 
model with such low scale dynamics where not only
the supersymmetry breaking but also the masses of the messenger particles 
are generated by the effects of the strong dynamics of a gauge interaction.

%%%%%%%%%%%%%%%%%%%%%%
\subsection{Dynamical supersymmetry breaking/messenger sector}
A model of the dynamics of the supersymmetry breaking
is based on a supersymmetric SU(5)$_{H}$ gauge theory with five flavors
($F$ and $\bar F$)~\cite{Izawa:1995jg},
where the subgroup of  a global SU(5)$_{F}$ symmetry 
($F:5$ and $\bar F: \bar 5$) is identified with the gauge groups of the standard model.
As we see later, dynamical supersymmetry breaking is therefore directly
 communicated to the standard model sector directly~\cite{Dine:1981za,Affleck:1984xz}.
The only other ingredient of the model is a singlet superfield $S$ which couples to $F$ and $\bar F$
in the superpotential,
\begin{eqnarray}
W = k S F\bar F,
\end{eqnarray}
where $k$ is a coupling constant.
$F$ and $\bar F$ (or their composite particles) serve as the messenger fields 
once both the scalar- and the $F$-term of the singlet obtain non-vanishing
vacuum expectation values; $\vev S \neq 0$ and $F_{S}\neq 0$
\footnote{
We use the same notation for chiral superfields and their scalar parts interchangeably.}.
Recently, models of {\it the sweet spot supersymmetry}~\cite{Ibe:2007km} based on this 
dynamical supersymmetry breaking/messenger model
have been analyzed in the elementary messenger regime~\cite{Nomura:2007cc}  
and in the composite messenger regime~\cite{Ibe:2007gf}. 
In {\it the sweet spot supersymmetry}, 
the supersymmetry breaking local minimum at $\vev S\neq 0$ is realized by
the gravitational stabilization mechanism~\cite{Kitano:2006wz}.
In this letter, we seek another possibility of making $\vev S\neq 0$.

To see how the supersymmetry breaking occurs, 
let us consider the region where the ``messenger mass'', $M_{\rm mess} \equiv k S$, 
is smaller than the scale $\Lambda$ around which the SU(5)$_{H}$ gauge interaction becomes strong.
In this region, the model can be described by using mesons,
$M\sim F\bar F$ and baryons $B\sim F^{5}$ and $\bar B\sim \bar F^{5}$.
Here, we omit the indices of the flavor SU(5)$_{F}$ symmetry.
In terms of the mesons and baryons, the full effective superpotential is given by
\begin{eqnarray}
\label{eq:Weff1}
W_{\rm eff}  = k S\cdot {\rm Tr} M + X (\det M -B \bar B - (\Lambda_{\rm dyn}^{2}/5)^{5}),
\end{eqnarray}
where $X$ is a Lagrange multiplier field which guarantees the quantum modified constraint between 
the mesons and baryons~\cite{Intriligator:1995ne}.
$\Lambda_{\rm dyn}$ denotes the dynamical scale of SU(5)$_{H}$ gauge interaction and it is naively related to the scale 
$\Lambda$ by $\Lambda_{\rm dyn}\sim\sqrt{N_{c}}\Lambda/(4\pi)$ 
with $N_{c}=5$~\cite{Luty:1997fk,Cohen:1997rt}.
By expanding the meson and baryon fields around a solution to the above constraint:
$M=\Lambda_{\rm dyn} (\Lambda_{\rm dyn}\delta_{ij}/5+ \delta \hat M$), (Tr$\delta \hat M$=0),
 $\hat B \sim B/\Lambda_{\rm dyn}^{4}$ and 
$\hat {\bar B} \sim \bar B/\Lambda_{\rm dyn}^{4}$,
we obtain a superpotential, 
\begin{eqnarray}
\label{eq:Weff2}
W_{\rm eff} \sim k \Lambda_{\rm dyn}^{2} S + S \left( \frac{k}{2} {\rm Tr} \delta \hat M^{2} 
+ k \hat B \hat{\bar B}
+ \cdots 
\right),
\end{eqnarray}
which has a linear term of the singlet $S$.
Here, the ellipses denote the higher dimensional operators of mesons and baryons, and
we have neglected non-calculable corrections to the coupling constants 
which are naively expected to be $O(1)$.
This superpotential shows that there is a supersymmetric minimum at $ S = 0$
and $\delta \hat M^{2} \neq 0$ or $\hat B\hat{\bar B} \neq 0$.
However, if the singlet $S$ has a local minimum at $S =\vev{S} \agt \Lambda_{\rm dyn}$, 
the mesons and baryons have positive masses squared and the spontaneous supersymmetry breaking is achieved 
by $F_{S} \sim k \Lambda_{\rm dyn}^{2}$.

Now, the question is: 
is there a possibility for the singlet $S$ to have a local minimum at $S=\vev S> \Lambda_{\rm dyn}$?
To address this question, notice that 
 there are non-calculable contributions to the effective K\"ahler potential of the singlet $S$
 from the strong interactions below the scale $\Lambda$,
\begin{eqnarray}
K_{\rm eff} = S^{\dagger} S 
+ \frac{25\Lambda^{2}}{(4\pi)^{2}} \delta K\left(\frac{k S}{\Lambda}
\right). 
\end{eqnarray}
Here, we have used the ``naive dimensional analysis''~\cite{Luty:1997fk,Cohen:1997rt}, 
and we expect that the non-calculable contribution $\delta K(x)$ has no small parameter.
From this K\"ahler potential, we obtain a scalar potential,
\begin{eqnarray}
\label{eq:scalarpot}
V(S) \sim \frac{|F_{S}|^{2}}{1 + 25(k/4\pi)^{2} \delta K^{(2)} (kS/\Lambda)},
\end{eqnarray}
where $\delta K^{(2)}$ denotes the second derivative of $\delta K$ with respect to 
$k S/\Lambda$.
This potential shows that there is a possibility that the potential has a local minimum around 
$\vev S \sim \Lambda/k $ which is larger than $\Lambda_{\rm dyn}$ for $k \alt 4\pi/\sqrt 5$.
Therefore, if we take this possibility positively, it is not hopeless to expect
that the singlet $S$ has a local minimum around $\Lambda/k$
by the effect of the non-calculable contributions from the strong dynamics  (see Fig.~1).

Note that calculable radiative corrections to the K\"ahler potential
through the diagrams in which the mesons and baryons circulate 
dominate over the non-caluculable contribution $\delta K$ for a small value of the singlet $S$,
$k S \ll \Lambda$. 
The potential curves up by this contribution
\footnote{
To ensure the origin of the singlet is not a local minimum in models with
$R$-symmetry like our model, 
we need to introduce fields with $R$-charges other than $0$ or $2$~\cite{Shih:2007av},
although we do not require the instability of the origin of the singlet in our model.
}.
The masses of the mesons and baryons, however,  become comparable to the scale $\Lambda$
around $S\sim \Lambda/k$.
Hence, their effects can be overwhelmed by the non-calculable contribution in the region of 
$S \sim \Lambda / k$.
Therefore, the possibility of the local minimum around $S \sim \Lambda/k$ can not 
be excluded by these effects (see Ref.~\cite{Chacko:1998si} for a similar discussion).

For $M_{\rm mess} = kS\gg  \Lambda$, the dynamics can be described by
using $F$  and $\bar F$ as elementary fields.
In this region, it can be shown that 
the potential curves up in the $S$ direction by radiative corrections to the Kahler potential~\cite{ArkaniHamed:1997ut}.
Therefore, the singlet $S$ cannot have a local minimum at $S \gg \Lambda / k$.

The above discussion does not exclude the possibility of having a local minimum around 
$S \sim \Lambda / k$ where all the calculable contributions are comparable to the non-calculable ones.
Put it all together positively, we here assume 
that there is a local and supersymmetry breaking minimum around 
$\vev S \sim \Lambda/k$,
aside from the supersymmetric minimum at $S = 0$
\footnote{
Here, we are also assuming that the K\"ahler metrics
of $\delta \hat M$, $\hat B$ and $\hat{\bar B}$ are positive definite
around the local minimum, 
although the composite description is not quite well for $k\vev S \sim  \Lambda$.}.
In the following, we consider a model with gauge mediation around the local minimum
at
\begin{eqnarray}
\label{eq:minS}
\vev S &\sim& \Lambda/k\, \sim\,
4\pi\Lambda_{\rm dyn}/(\sqrt{5}k )
\,\agt\, \Lambda_{\rm dyn},\\
F_{S} &\sim& k \Lambda_{\rm dyn}^{2},\\
\label{eq:minH}
\vev{\delta \hat M}&=&\vev{\hat B}\, = \,\vev{\hat{\bar B}} \, =\,0.
\end{eqnarray}

%%%%%%%%%%%%%%%%%%%%%%%%%%%%%%%%%%%%%%%%%%%%%
\begin{figure}[t]
\begin{center}
  \includegraphics[width=.8\linewidth]{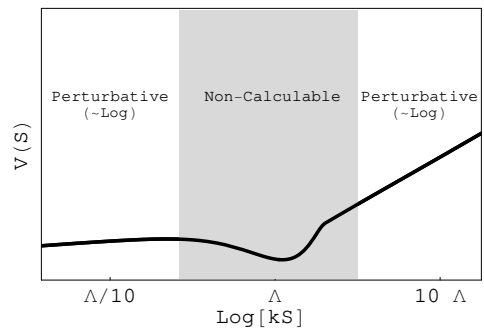}
  \end{center}
\caption{Schematic picture of the scalar potential $V(S)$.
In the dark shaded region, there is a possibility that $V(S)$ has a local minimum
due to the non-caluculable contribution to the K\"ahler potential from the strong dynamics.
}
\label{fig:gravitino}
\end{figure}
%%%%%%%%%%%%%%%%%%%%%%%%%%%%%%%%%%%%%%%%%%%%%

%%%%%%%%%%%%%%%%%%%%%%%
\subsection{Spectrum of supersymmetric standard model particles}
The mesons $\delta \hat M$ transform as the adjoint representation under SU(5)$_{F}$
whose subgroup is identified with the standard model gauge group,
(i.e. $(8,1)_{0}$, $(1,3)_{0}$, $(3,2)_{-5/6}$ and $(\bar 3, \bar 2)_{5/6}$ under the standard group).
Thus, the mesons mediate the effects of the supersymmetry breaking to the standard model sector
via gauge interactions of the standard model.
Through loop diagrams of the mesons,
we obtain masses of the  gauginos and scalar particles in the standard model sector,
\begin{eqnarray}
m_{\rm gaugino} &=& N_{m}\frac{g^{2} }{(4\pi)^{2}} 
\frac{F_{S}}{\vev{S}}
\left(
1 + O\left( ( 
\sqrt 5k/4\pi)^{4}\right)
\right),
\\
m_{\rm scalar}^{2} &=& 2N_{m}C_{2}\eta\left(\frac{g^{2}}{(4\pi)^{2}} \right)^{2}
\left|\frac{F_{S}}{\vev{S}}\right|^{2},
\end{eqnarray}
where $N_{m} = 5$ is the sum of the Dynkin index of the mesons, $C_{2}$ is
the quadratic Casmir invariant of the scalar particles,  and $g$ denotes
the gauge coupling constant of the standard model gauge group.
As we discussed above, $F_{S}/\vev S$ is given by 
$F_{S}/\vev S \sim \sqrt{5}k^{2} \Lambda_{\rm dyn}/4\pi$.
The $O(1)$ coefficient $\eta$ for the scalar masses denotes non-caluculable 
$O(F_{S}^{2}/\vev S^{2})$ contributions from the heavy hadrons which are charged under the standard model gauge groups, while gaugino masses do not get $O(F_{S}/\vev S)$ contributions from them.
The $O(( \sqrt 5k/4\pi)^{4})$ contribution comes from the diagrams with more $S$ inserted.
As a result, we achieve a model with gauge mediation with a low dynamical scale,
\begin{eqnarray}
\label{eq:paramL}
\Lambda_{\rm dyn} \sim  10^{5}\,{\rm GeV}\times k^{-2} \left(\frac{m_{\rm gluino}}{1\,{\rm TeV}} \right),
\end{eqnarray}
for $k=O(1)$.

As an interesting prediction, the gravitino mass can be as small as $O(1)$\,eV,
\begin{eqnarray}
m_{3/2}\sim 3 \,{\rm eV}\times k^{-3} \left(\frac{m_{\rm gluino}}{1\,{\rm TeV}}\right)^{2},
\end{eqnarray}
for $\Lambda_{\rm dyn}\sim 10^{5}$\,GeV%
\footnote{
With such a small gravitino mass, it has been known that the overproduction problem 
by thermal scattering processes is absent~\cite{Viel:2005qj}. 
However, in this model in order to avoid destabilization of the meta-stable vacuum by thermal effects, 
we need to assume a low reheating temperature and/or a low scale inflation. 
We thank T.~Yanagida for pointing this out.
}.
Here, we have used the definition of the gravitino mass,
\begin{eqnarray}
m_{3/2} = \frac{F_{S}}{\sqrt{3}M_{\rm pl}} \sim \frac{k \Lambda_{\rm dyn}^{2}}{\sqrt{3}M_{\rm pl}},
\end{eqnarray}
where $M_{\rm pl}\simeq 2.4 \times 10^{18}$\,GeV denotes the reduced Planck scale.

Now several comments are in order.
The perturbativity of the standard model gauge interactions up to the scale of the Grand Unification Theory (GUT)
put a bound on the sum of the Dynkin index of the messenger field $N_{m}$ as 
\begin{eqnarray}
 N_{m} \alt 150/ \ln(M_{\rm GUT}/M_{\rm mess}),
\end{eqnarray}
where $M_{\rm GUT}\simeq 2\times 10^{16}$\,GeV denotes the scale of GUT.
For $M_{\rm mess}\sim\Lambda\sim 10^{6}$\,GeV,
this condition requires $N_{m}$  as $N_{m} \leq 6$.
Therefore, the Dynkin index of the present model, $N_{m}=5$, 
satisfies the perturbative condition of the standard model gauge interactions up to the GUT scale.%

We should also mention the perturbativity of the coupling constant $k$.
The coupling constant $k$ becomes small at the high energy scale as a result of the large renormalization effect from the strong gauge interaction of SU(5)$_{H}$.
Thus, we can expect the coupling constant $k$ stays perturbative up to around the GUT scale,
although it is not necessarily required
\footnote{
We cannot trace the renormalization group evolution
from the value in Eq.~(\ref{eq:paramL}) since
there is non-calculable $O(1)$ threshold corrections 
between the value of $k$ at the scale $\Lambda$
and the one at the scale higher than $ \Lambda$.}.

The tunneling rate 
to the supersymmetric vacuum at $S = 0$ per unit volume  is
roughly given by $\Gamma/V \sim \vev S^{4}  e^{-{\cal S}_{ E}}$,
where ${\cal S}_{ E}$ is estimated by
${\cal S}_{ E} \sim 2\pi^{2}\vev{S}^{4}/V(S) \sim  2 \pi^{2}(4\pi)^{4}/(5^{2}k^{6}) \sim 10^{4}$ for 
$k\sim 1$.
On the other hand, the vacuum stability condition within the observable volume and
over the age of the universe only requires ${\cal S}_{E}\agt  400$.
Thus, although our vacuum is not stable quantum mechanically, 
it has a lifetime much longer than the age of the universe.

Finally, we comment on the effects of the supergravity to the scalar potential of $S$.
The leading effect of the supergravity comes from the linear term of the singlet $S$ in the superpotential
which leads to a linear term in the scalar potential,
\begin{eqnarray}
V(S)_{\rm linear} = 2 m_{3/2} k  \Lambda_{\rm dyn}^{2} S + h.c.
\end{eqnarray}
The linear term, however, is negligible compared with the scalar potential in Eq.~(\ref{eq:scalarpot})
around $S\sim \Lambda/k$ as long as,
\begin{eqnarray}
\label{eq:gravitational}
k \agt 4 \left( \frac{\Lambda_{\rm dyn}}{M_{\rm pl}} \right)^{1/3}. 
\end{eqnarray}
Here we have used $\partial V(S)/\partial S \sim 5 (\sqrt{5}k/4\pi)^{3} |F_{S}|^{2}/\Lambda_{\rm dyn}$ for
$S \sim 4\pi\Lambda_{\rm dyn}/(\sqrt 5 k)$.
The condition is easily satisfied for $\Lambda_{\rm dyn}\sim 10^{5}$\,GeV ($k = O(1)$) 
(Eq.~({\ref{eq:paramL}})),
and hence, the local minimum we chose is stable against the supergravity effects.

On the other hand, the linear term plays an important role to generate 
the mass of the $R$-axion which corresponds to the spontaneous breaking 
of the $R$-symmetry by $S\neq 0$.
Since the $R$-symmetry is broken explicitly by the linear term,
the $R$-axion obtains a mass~\cite{Bagger:1994hh},
\begin{eqnarray}
 m_{a} &\simeq &2 m_{3/2} \left(\frac{M_{\rm pl}}{\vev S}\right)^{1/2} \\
 \nonumber
& \sim &10\,{\rm MeV}  \times k^{-3/2}
  \left( \frac{m_{\rm gluino}}{1\,{\rm TeV}} \right)^{3/2}.
\end{eqnarray}
The $R$-axion couples to the standard model particles through the loop
diagrams of mesons which are relevant for the gauge mediation. 
As a result, it decays mainly into photons at the cosmic temperature $T\sim O(10)$\,MeV. 
Note that for the axion with mass $m_{a} \sim O(10)$\,MeV, 
final states with hadrons or electroweak gauge bosons are kinematically forbidden. 
The cosmic abundance of the $R$-axion before the decay 
(both from thermal and non-thermal production)
is estimated to be small enough that the decay does not cause a large entropy production
\footnote{The decay mode into gravitinos are suppressed by a helicity suppression
factor of $(m_{3/2}/m_{a})^{2}$, and hence, the gravitino abundance produced 
by the decay of the $R$-axion is negligible compared with the one
from the thermal bath~\cite{Bagger:1994hh}.}.
The decay temperature is also high enough not to spoil the success of the Big-Bang Nucleosynthesis.
Besides, the above $R$-axion marginally satisfies an 
 astrophysical constraint based on stellar cooling rate and supernova dynamics: 
 $m_{a}\agt O(10)$\,MeV~\cite{Raffelt:1990yz}. 
Therefore, we find that the $R$-axion in our model does not cause any cosmological 
and astrophysical problems.

%%%%%%%%%%%%%%%%%%%%%%%%%%%%
\subsection{Conclusions}
We find a very simple model with gauge mediation where the supersymmetry breaking/mediation
is realized by a dynamics at $\Lambda_{\rm dyn}\sim 10^{5}$\,GeV.
Furthermore, the model predicts a very small gravitino mass ($m_{3/2}\sim O(1)$\,eV) 
for $\Lambda_{\rm dyn}\sim 10^{5}$\,GeV,
which can be measured at the future collider experiments such as LHC/ILC, 
by measuring the branching ratio of the decay rate of the next to lightest 
superparticle~\cite{Hamaguchi:2007ge}.

Finally, it should be noted that the present model is also applicable for a wide range of the dynamical scale up to $\Lambda_{\rm dyn}\sim 10^{10}$\,GeV, ($k\sim 10^{-2}$, $m_{3/2}\sim O(10)$\,MeV), 
where the condition in Eq.~(\ref{eq:gravitational}) is saturated.

\section*{Acknowledgments}
MI thanks Y.~Nakayama and T.T.~Yanagida for useful discussion.
The work of MI was supported by the U.S. Department of Energy under contract number 
DE-AC02-76SF00515.

%\newpage %Just because of unusual number of tables stacked at end
%\bibliography{file}% Produces the bibliography via BibTeX (file.bib).

\end{document}